# Superconductivity in Epitaxial Thin Films of Co-Doped SrFe$_2$As$_2$ with Bilayered FeAs Structures and their Magnetic Anisotropy

Hidenori Hiramatsu[1,*], Takayoshi Katase[2], Toshio Kamiya[1,2], Masahiro Hirano[1,3], and Hideo Hosono[1,2,3]

[1]*ERATO–SORST, Japan Science and Technology Agency (JST), in Frontier Research Center, Tokyo Institute of Technology, S2-6F East, Mail-box S2-13, 4259 Nagatsuta-cho, Midori-ku, Yokohama 226-8503, Japan*

[2]*Materials and Structures Laboratory, Mail-box R3-1, Tokyo Institute of Technology, 4259 Nagatsuta-cho, Midori-ku, Yokohama 226-8503, Japan*

[3]*Frontier Research Center, S2-6F East, Mail-box S2-13, Tokyo Institute of Technology, 4259 Nagatsuta-cho, Midori-ku, Yokohama 226-8503, Japan*

**ABSTRACT**

Superconducting epitaxial films of Fe-based layered arsenide, Co-doped SrFe$_2$As$_2$, were grown at 700°C on mixed perovskite (La, Sr)(Al, Ta)O$_3$ (001) single-crystal substrates by pulsed-laser deposition. Both the epitaxial film and an (001)-oriented film grown at 600°C exhibited superconducting transitions at ~ 20 K. The zero-resistance states of the epitaxial film were sustained under a magnetic field ($H$) of 9 T at 9 K when $H$ was parallel to the $c$-axis, while they were sustained at higher temperatures up to 10 K for $H$ parallel to the $a$-axis. This is the first demonstration of superconducting thin films of FeAs-based new superconductors.

[*]E-mail address: h-hirama@lucid.msl.titech.ac.jp

Since the recent discovery of superconduction in F-doped LaFeAsO [1, 2], extensive studies on Fe-based layered oxypnictides $Ln$Fe$Pn$O ($Ln$ = lanthanides, $Pn$ = pnicogens) have been conducted, and the transition temperature ($T_c$) has reached 55 K in SmFeAsO [3–5]. In addition, superconductivity has been found for relevant compounds having bilayer Fe$Pn$ structures, $AE$Fe$_2$$Pn$$_2$ ($AE$ = alkaline earth elements) [6–8], with appropriate substitutional doping such as K-doping at the $AE$ site. As such, a variety of superconductors has been found for Fe-based layered compounds in the last seven months. Much progress has been made in the fundamental physics of these materials as well, i.e., their carrier transport, ground state charge/spin orderings, and electronic structure have been elucidated both from experimental and theoretical aspects. However, there are still controversies regarding the fundamental properties. The primary origin of this controversy is that most of the materials' properties reported to date were measured on bulk polycrystals, except for some single-crystal works [9–19]. Therefore, high-quality epitaxial films are required to explore intrinsic properties and to develop superconductor electronic devices such as the Josephson device.

Thin films of Fe-based layered compounds have not yet been reported even for a polycrystalline film. Very recently, we succeeded in growing epitaxial films of F-doped LaFeAsO [20], but they did not exhibit a superconducting transition. In that work, we recognized that the difficulty of thin film growth of $Ln$Fe$Pn$O may come from the crystal structure containing two different anions, because it is difficult to control their stoichiometry, especially for thin film growth. In addition, $Ln$Fe$Pn$O needs electron doping, e.g., by partially replacing the oxygen ions with F to induce a superconducting transition, but such dopants are easily evaporated in a vacuum chamber at a high temperature. Recently, it was reported that Co doping at the Fe site induces superconducting transitions in LaFeAsO [21, 22] and $AE$Fe$_2$As$_2$ [23, 24]. Although the reported $T_c$ values are lower than those obtained by substitution of the O or $AE$ site, we expect that Co doping is more suitable for thin film growth than F and K doping because the vapor pressures of Co-related species are much lower than those of the F and K compounds. In addition, SrFe$_2$As$_2$ contains only one anion species and, therefore, is expected to grow thin films more easily than mixed-anion compounds such as *Ln*Fe*Pn*O.

In this study, we report epitaxial film growth of Co-doped SrFe$_2$As$_2$ and its electronic transport properties under a magnetic field. Epitaxial films have been obtained at 700°C, while (001)-oriented films have been obtained at 600°C by pulsed-laser deposition (PLD). The resulting epitaxial films exhibited superconducting transitions at ~ 20 K, and the superconducting phase was sustained under high magnetic fields at least up to 9 T at 9 K.

Co-doped SrFe$_2$As$_2$ thin films were fabricated on mixed perovskite (La, Sr)(Al, Ta)O$_3$ (LSAT) (001) single-crystal substrates by PLD using a second-harmonic Nd:YAG laser (wavelength: 532 nm, frequency: 10 Hz, fluence: ~1.5 J/cm$^2$) and a target disk of SrFe$_{1.8}$Co$_{0.2}$As$_2$. The synthesis condition of the target disk was followed according to ref. [23]. Substrate temperature ($T_s$) was varied from 500 to 870°C, and the deposition atmosphere was in a vacuum at ~10$^{-5}$ Pa. The film thicknesses were ~350 nm, which were measured with a surface profiler.

Crystalline quality and orientation were characterized with a high-resolution X-ray diffraction (HR-XRD) apparatus (ATX-G, RIGAKU Co., anode power: 50 kV × 300 mA, radiation: Cu K$\alpha_1$) at room temperature (RT). Out-of-plane ($\omega$-coupled $2\theta$ scan) and in-plane ($\phi$-coupled $2\theta_\chi$ scan) HR-XRD patterns were measured to determine epitaxial relationships between the films and the LSAT substrates. Rocking curves of out-of-plane ($2\theta$-fixed $\omega$ scans) and in-plane ($2\theta_\chi$-fixed $\phi$ scans) were measured to evaluate fluctuations of the crystallites orientation in the films. Electrical resistivity measurements were performed by a four-probe method in the temperature range of 2–305 K using a physical property measurement system (PPMS, Quantum Design). Magnetic fields ($H$) of 0–9 T were applied in the resistivity measurements normal to and parallel to the film surfaces.

We first examined the effects of $T_s$ on the film structure. At $T_s$ < 600°C, the SrFe$_2$As$_2$ phase was not detected by HR-XRD. The formation of SrFe$_2$As$_2$ films was observed at higher $T_s$ along with an impurity phase of FeAs, but a further increase in $T_s$ to ≥750°C broadened the diffraction peaks, accompanied by an increase in the FeAs impurity content. At $T_s \geq$ 800°C, only the FeAs phase remained. Therefore, we will describe the results of the films grown at 600 and 700°C hereafter.

Figure 1 shows the out-of-plane (a) and in-plane (b) HR-XRD patterns of the Co-doped SrFe$_2$As$_2$ thin films grown at 600°C (bottom panels) and 700°C (top panels). The out-of-plane HR-XRD pattern of the 600°C film exhibits sharp SrFe$_2$As$_2$ 00$l$ diffraction peaks along with the LSAT 004 diffraction, indicating that the film is preferentially oriented along the $c$-axis [bottom panel of Fig. 1(a)]. The full width at half maximum (FWHM) of the 002 rocking curve was as large as 4 deg. Clear in-plane diffraction from the SrFe$_2$As$_2$ phase was not observed, and a small amount of the FeAs phase was detected in the in-plane HR-XRD pattern [bottom in Fig. 1(b)]. These results indicate that the 600°C film constituted mainly of SrFe$_2$As$_2$ with a small amount of FeAs. The SrFe$_2$As$_2$ phase was weakly oriented along the $c$-axis, but in-plane orientation was not determined.

When $T_s$ was raised to 700°C (upper panels in Fig. 1), the FWHM of the 002 rocking curve became sharper to 1 deg, and the 200 in-plane diffraction of SrFe$_2$As$_2$ was clearly observed along with the 400 diffraction of LSAT, although the diffraction intensity from the FeAs impurity increased. The in-plane rocking curve ($\phi$ scan) of the 200 diffraction [inset of top panel in Fig. 1(b)] shows that the 700°C film has a four-fold in-plane orientation, which corresponds to the tetragonal symmetry of the SrFe$_2$As$_2$ lattice [23]. These observations substantiate that the 700°C film was grown heteroepitaxially on the LSAT (001) single-crystal substrate with the epitaxial relationship of (001)[100] SrFe$_2$As$_2$ ∥ (001)[100] LSAT.

From the microstructural ordering structures observed by HR-XRD, we will hereafter refer to the films grown at 600 and 700°C as low-quality (LQ) and high-quality (HQ) films, respectively. The lattice parameters of the HQ film were $a$ = 0.3936 nm and $c$ = 1.231 nm, which are close to those of a bulk sample with the same chemical composition, $SrFe_{1.8}Co_{0.2}As_2$ [$a$ = 0.39278(2) nm and $c$ = 1.23026(2) nm] [23], but slightly larger (~0.2% for the $a$-axis, ~0.06% for the $c$-axis). Both the HQ and LQ films exhibit metallic glossy appearances if the reflection light is observed [inset of bottom panel in Fig. 1(b)], while these appear black when viewed from the direction normal to the film surfaces. The black color is the same as that observed in the powder and polycrystalline bulk samples and compatible with a zero or a very small band gap of Co-doped $SrFe_2As_2$. The metallic glossy is consistent with metallic conduction, which will be shown in the next paragraph.

The LQ film showed almost the same resistivity (4.7 × 10$^{-4}$ Ωcm at RT) as those of bulk samples of Fe-based arsenides with a bilayered FeAs structure [6–8, 24], while the HQ film was more conductive with smaller resistivities of 3.6 × 10$^{-4}$ Ωcm (inset of Fig. 2), suggesting that the crystallinity and the carrier transport of the HQ film are better than those of the LQ film and the bulk samples. The resistivity of both the samples started to drop sharply from 20 K, and zero resistivity was observed at <14–16 K for the HQ and LQ films (Fig. 2). This result implies that the obtained films are superconductors with the onset $T_c$ ~ 20 K. The resistivity drop was sharper for the LQ film than for the HQ film notwithstanding that the crystallinity assessed by the HR-XRD measurements was better for the HQ film. The $T_c$ values were almost the same as that of the $SrFe_{1.8}Co_{0.2}As_2$ polycrystalline bulk samples in ref. [23], suggesting that the chemical composition of the PLD target ($SrFe_{1.8}Co_{0.2}As_2$), including the doping concentration of Co, was transferred almost directly to the films.

Figure 3 shows the effects of external magnetic fields ($H$) on the superconducting transitions of the HQ (top panels) and LQ (bottom panels) films. The magnetic fields were applied parallel to the $c$-axis (a) and to the $a$-axis (b) of the LSAT substrates, where the latter direction corresponds to $SrFe_2As_2$ [100] for the HQ film. However, the in-plane crystallographic direction cannot be determined for the LQ film because it did not show in-plane orientation in the HR-XRD pattern. It is observed that $T_c$ was shifted to lower temperatures as $H$ increased. It should be noted that the superconduction was observed in both the films even when $H$ reached 9 T at ≤7 K, indicating that the upper critical magnetic fields ($H_{c2}$) are much higher than 9 T [14, 25, 26]. The superconducting states are more sensitive to the magnetic fields when $H$ is parallel to the $c$-axis than to the $a$-axis, but the difference is not large despite the layered crystal structure [26]. It would also be worth noting that the slope of the resistivity drop below $T_c$ became smaller with increasing $H$ for the LQ film, while the slope for the HQ film did not exhibit a large change. Consequently, the HQ film shows zero resistivity at higher temperatures [≤10 K ($H // a$) and ≤9 K ($H // c$) at 9 T] compared to the LQ film [≤9.5 K ($H // a$-$b$ plane) and ≤7 K ($H // c$) at 9 T] under strong magnetic fields.

In summary, we have succeeded in fabricating epitaxial films of an Fe-based layered superconductor, Co-doped $SrFe_2As_2$. The epitaxial films exhibited $T_c$ ~ 20 K, which is the same as that of the polycrystalline bulk samples, and very high upper critical magnetic fields >9 T at ≤9 K. The anisotropy of the magnetic field effects was also observed, but it was found not to be large in $SrFe_2As_2$. It is expected that epitaxial films of the new superconductors will contribute to measuring the intrinsic properties and to clarifying the mechanisms of superconduction common to these materials. We would also like to point out that we have a large flexibility in choosing appropriate electrical properties in pseudo-isostructural crystals: Fe-based layered compounds show metallic states for undoped states, and the pseudo-isostructural crystals of LaFeAsO include wide band gap insulators (e.g., $La_2CdSe_2O_2$ [27, 28]), antiferromagnetic semiconductors (e.g., LaMn$Pn$O [29, 30]), wide band gap p-type semiconductors (e.g., LaCuSeO [31–33]), and so on. We have developed epitaxial growth techniques for these materials, which will lead to the development of superconductor devices based on pseudo-homojunctions of an Fe-based layered superconductor and a pseudo-isostructural insulator/metal.

We would like to thank Dr. Satoru Matsuishi of Frontier Research Center, Tokyo Institute of

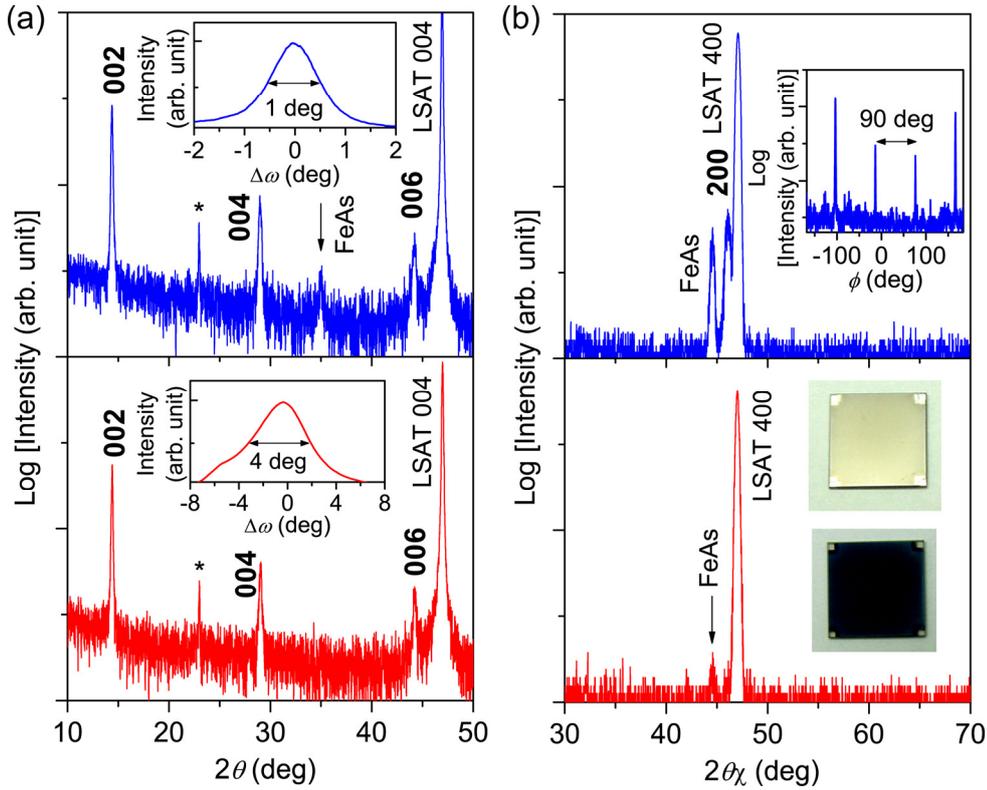

Fig. 1. Out-of-plane (a) and in-plane (b) HR-XRD patterns of the Co-doped $SrFe_2As_2$ thin films grown at 700°C (top panels, HQ: high-quality) and 600°C (bottom panels, LQ: low-quality). Asterisks in (a) indicate the LSAT 002 diffraction that is an extinct diffraction in the kinematical diffraction theory but appears due to multiple scattering. Insets in (a) show the out-of-plane rocking curves ($\omega$ scans) of the 002 diffraction. Inset in the top panel of (b) shows the in-plane rocking curve ($\phi$ scan) of the 200 diffraction. In-plane diffraction from the $SrFe_2As_2$ phase was not observed for the LQ film. Inset in the bottom panel of (b) shows photographs of the HQ film (size: $1 \times 1$ cm$^2$). The film exhibits metallic glossy appearances as seen in the upper photograph, but the simple top view (the bottom photograph) shows black color.

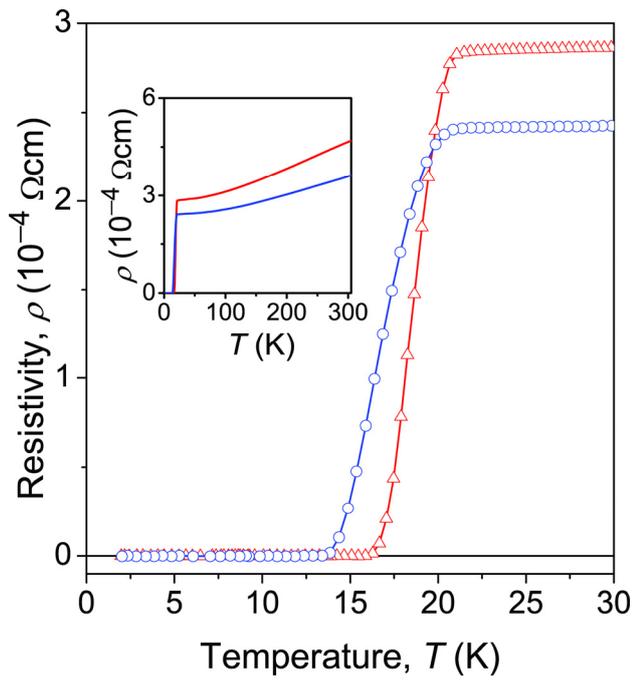

Fig. 2. Temperature dependencies of electrical resistivities of the Co-doped $SrFe_2As_2$ thin films. Inset shows expanded views between 2 and 305 K. Blue circles and red triangles show the HQ and LQ films, respectively.

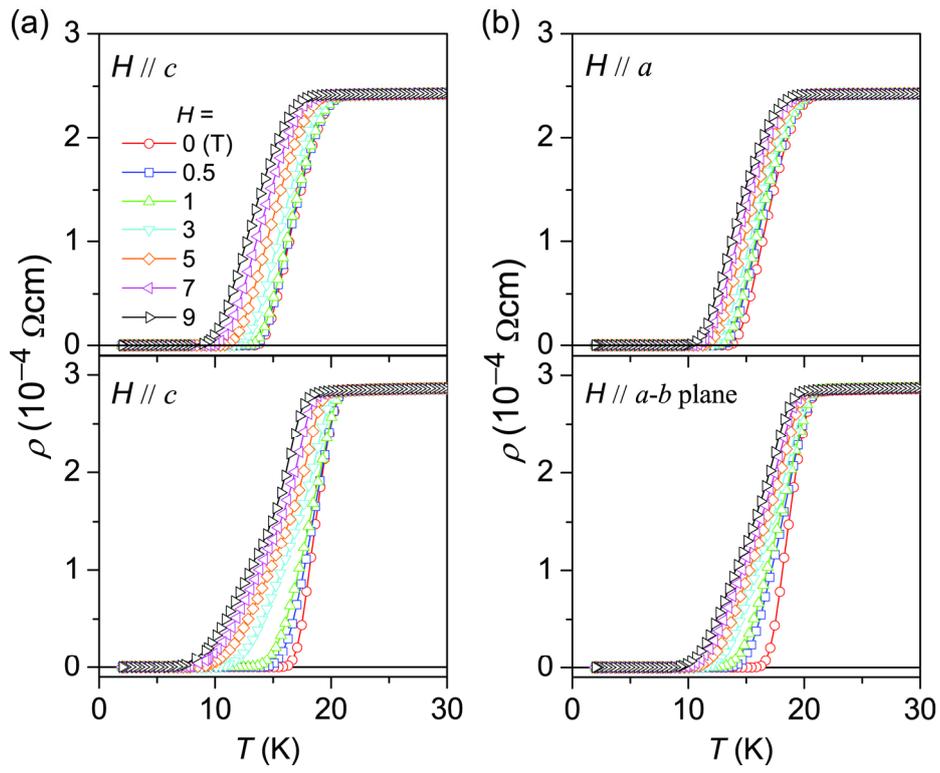

Fig. 3. Superconducting transitions under magnetic fields (*H*) varying from 0 to 9 T applied parallel to the *c*-axis (a) and the *a*-axis of the LSAT substrates (b). The top and bottom panels correspond to the HQ and LQ films, respectively.